# Development of simulation package for atomic processes of ultra-large-scale system based on electronic structure theory

Hitoshi Nitta[1], Naoki Watanabe[1], Takeo Hoshi[2, 3] and Takeo Fujiwara[3, 4,5]

[1]Science Solutions Division, Mizuho Information & Research Institute, Tokyo, Japan; [2]Department of Applied Mathematics and Physics, Tottori University, Tottori Japan; [3]Core Research for Evolutional Science and Technology, Japan Science and Technology Agency, Saitama, Japan; [4]Center for Research and Development of Higher Education, University of Tokyo, Tokyo, Japan; [5]Department of Applied Physics, University of Tokyo, Tokyo, Japan.

## Abstract

An early-stage version of simulation package is developed for electronic structure calculation and dynamics of atom process in large-scale systems, particularly, nm-scale or 10nm-scale systems. We adopted the Extensible Markup Language (XML)-style in the input and the output of our simulation code, and developed some modeling and analysis tools for dynamical simulations of atomic processes. GaAs bulk system was calculated to demonstrate that the present code can handle systems with more than one atom specie.

## I. Introduction

A role of electronic structure calculation is to provide analysis or prediction tools of material structure and function within quantum mechanics. Nowadays, the first-principles molecular-dynamics simulations [1, 2], with pseudo-potential theory and plane-wave basis, is well established for system with $10^1$-$10^2$ atoms. However, these computational methods can be applied only small fraction of atomic system up to about a few angstrom such as bulk systems with periodic boundary condition and small molecules. Therefore, the reformative computational techniques and analysis algorithms have been required to solve the electronic structure and the dynamics of atomic processes occurring in the large scale atomic systems such as surface, amorphous, impurity-doped systems, and so on.

For years, some of the authors (T. Hoshi and T. Fujiwara) have developed a set of theories and program code for nanostructure process with electronic structure theory. [3, 4, 5, 6, 7, 8, 9, 10, 11, 12, 13,14] One crucial point is that large-scale quantum-mechanical calculation can be realized, in principal, by calculating the one-body density matrix, instead of one-electron eigen states, since the computational cost can be drastically reduced. [15] An overview of these theories can be found in the introduction part of Ref. [11] or review articles. [16, 17] Practical methods were constructed as solver methods of the one-body density matrix or the Green's function for a given Hamiltonian matrix. [3, 6, 7, 8, 10, 11, 12] We note that some of the theories are purely mathematical ones, iterative linear-algebraic algorithms for large matrices and, therefore, should be useful in other fields of physics. Actually, one method, called 'shifted conjugate-orthogonal conjugate-gradient method', [10] was applied to an extended Hubbard model for $La_{2-x}Sr_xNi_2O_4$. [18] Another crucial point is to construct algorithms for efficient parallel computations. Since multi-core CPU architectures are now built in standard workstation or personal computer, parallel computations are essential for actually all the computational systems. The calculations are realized with Slater-Koster-form (tight-binding) Hamiltonians and test calculation was carried out with $10^2$-$10^7$ atoms with or without parallelism. [6, 8, 11, 14] As a benchmark with a recent multi-core CPU architecture, we have tested our code with a standard workstation with four dual-core CPU's (Opteron 2GHz), for liquid carbon with 1728 atoms. We adopted a typical Hamiltonian of carbon system. [20] As a result, a computational time is six seconds per time step in the process (molecular dynamics) calculations and a parallel efficiency is more than 90%. We note the the electronic property, such as density of states, is also calculated. [7, 10, 11]

Now the code has named Extra Large Scale Electronic Structure calculation (ELSES) code, and it is being reorganized as a simulation package for a wider range of users and applications (www.elses.jp). Programing effort is being consumed mainly on the following three points; (1) Input and output files are formatted in a sophisticated style of Extensible Markup Language (XML). (2) Several tools are available that are used before and after the simulation. (3) The code can handle systems with more than one atom specie.

## II. Improvement of simulation package

  Figure 1 indicates the execution flow of our simulation code, showing input and output files and some examples of the calculated results. In order to run our simulation code, two input files formatted in the XML style are necessary to be prepared. One of these controls the calculation conditions such as solver methods of the one-body density matrix, dimension of Krylov space for the electronic structure calculation, time step of molecular dynamics calculations, and so on. Remaining one specify the initial atomic structure of molecular dynamics calculation. Figure 2 is an example of the XML-style input file, in which each input item is designated by start tag <...> and end tag </...>, and those are allowed to be freely defined by XML users. We note that the implementation of a XML-style input file is important in practical molecular dynamics simulations of nanostructure materials, since various conditions are required in these simulations. For example, the fracture simulation of silicon nanocrystal [8] was realized by imposing an external load on the atoms in a limited region neat the sample boundary. The extendibility of XML-style file can satisfy these detailed conditions, by adding newly-defined tags for its own purpose. The units of input quantities are selective owing to XML-style, so that the wide rage of users can choose their familiar units by setting as unit='fsec' for time, and unit='angstrom', 'nm', etc. for length quantities.

  Several pre- and post-processing tools are developed as independent tools and other tools will be developed according to user's or developer's demands. Since the data files are written in XML-style, these tools share the data-read or data-write subroutines and can be easily developed. By using pre-processing modeling tools, we can create the XML-style initial atomic structure file from the XYZ formatted one, that is widely used in the atomic simulation, and from output files of other atomic simulation codes. In order to investigate physical properties derived from calculated result, it is necessary to analyze the dynamics of atomic processes and electronic structure (density matrix and/or Green's function) in detail as well as to visualize atomic structures. The post-processing analysis tools may be useful to deal with a plenty of data resulted from simulations of large scale atomic systems, and gives meaningful physical quantities as shown in Figure 3. The tools should be developed further, particularly, for analyzing electronic structure. For example, a chemical bond between two atoms can be defined rigorously from electronic structure, not from the distance between the atoms, by calculating crystal orbital Hamiltonian populations (COHP) [19, 7] that is given the off-site elements of the Green's function. Moreover, the calculation code of transport properties, such as electronic current, should be also focused on.

## III. Application to compound; example of bulk GaAs

Bulk GaAs was calculated so as to demonstrate that the present code can handle systems with more than one atom specie. We adopt a Slater-Koster-form Hamiltonian of GaAs with s, p and s* atomic orbitals. [21] The atomic energy level of the s* orbital is located within the conduction band and its physical origin is a spherical average of the five d orbitals. The formulation of s, p, and s* orbitals was introduced [22] among various semiconductors, for reproducing the valence band and the bottom of the conduction band and was used in papers, such as Refs. [21, 23, 24, 25, 26, 27, 28], for liquid, amorphous, defect, surface and quantum dot.

Figure 4 shows calculation results of bulk GaAs, in which the cubic periodic cell with 64 atoms is used. Here the Krylov-subspace method with subspace diagonalization [7, 11] is adopted for solver routine of the density matrix. The dimension of the Krylov subspace (Krylov dimension) should be set as a controlling parameter that determines accuracy and computational cost. The computational time is proportional to the Krylov dimension and the calculation will be converged to the exact one, when the Krylov dimension increases. See Ref. [11] for detail. Figure 4 plots the optimized lattice constant and the energy as the function of the Krylov dimension. Figure 4 indicates that the calculation is well converged with the Krylov dimension of 30; the deviations in the lattice constant and the energy are less than 0.01% and less than 1meV per atom, respectively. We note that an excellent convergence at the Krylov dimension of 30 was found in the other systems. [7, 11]

## IV. Summary

Large-scale electronic structure calculation code is being developed as a simulation package with the name of ELSES (www.elses.jp). For a better user interface of our simulation code, we have created the input/output interfaces of XML-style files. The pre- and post-processing tools have been also prepared for modeling and detailed analysis of the atomic structures. We have also confirmed that the present code can handle system with more than one atom specie by calculating bulk GaAs. Although the present stage of the simulation package is still in an early one, we believe that the code will provide fruitful simulations for researchers in surface science and nano-material industry. More information may be found in the web page (www.elses.jp).

## Acknowledgements

Numerical calculation was partly carried out the Institute for Solid State Physics, University of Tokyo, and Research Center for Computational Science, Okazaki.

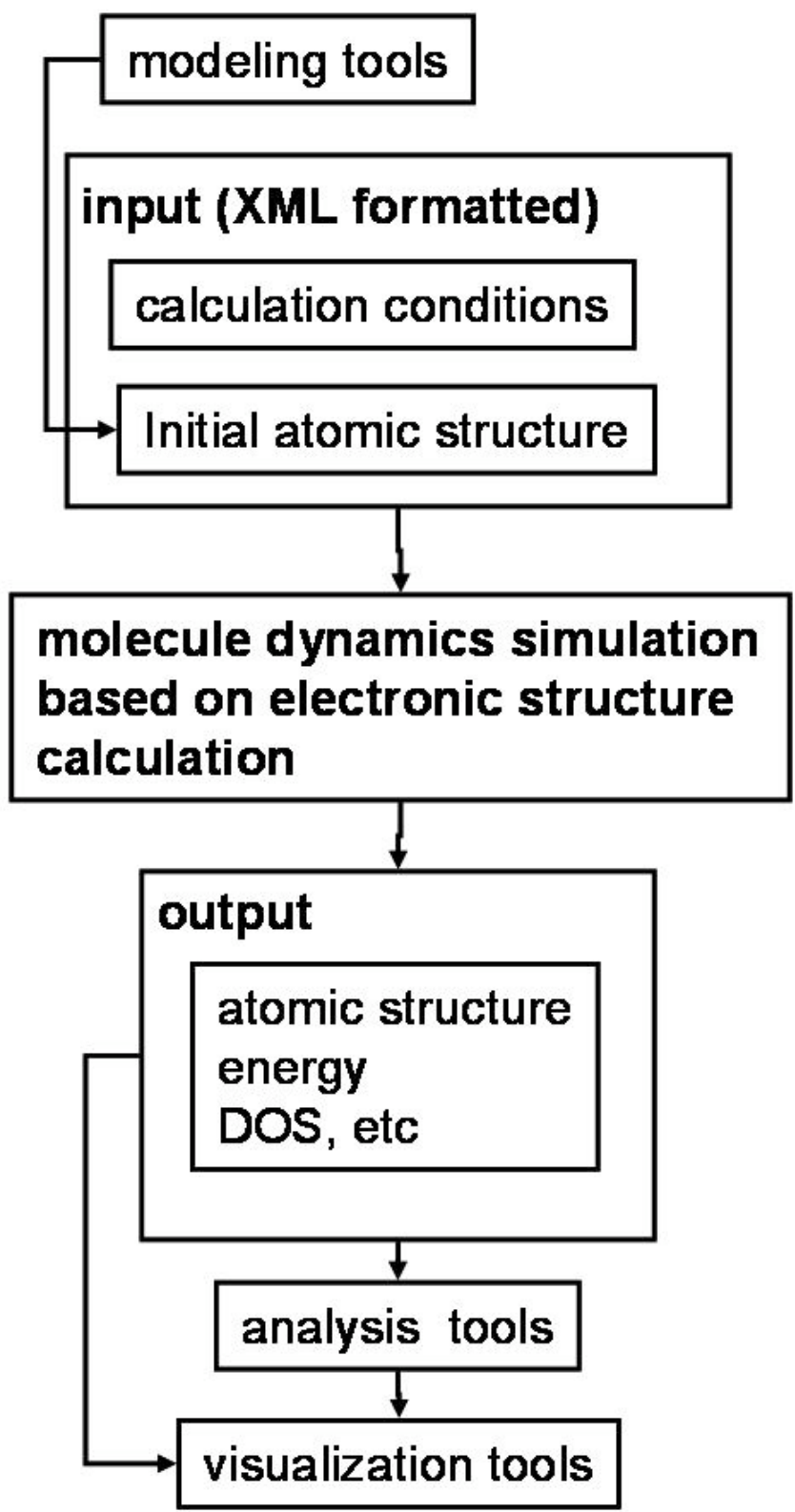

Figure 1 : Execution flow of simulation code; depicting input and output files and some examples of calculated results of our simulation..

```xml
<?xml version="1.0" encoding="UTF-8"?>
<config name="C60">

<system>
  <cluster structure="C60.xml" />

  <temperature unit="kelvin"> 273.15 </temperature>
</system>

<calc>
  <dynamics scheme="velocity verlet">
    <delta unit="fsec"> 1.00 </delta>
    <total unit="fsec"> 256.00 </total>
  </dynamics>
</calc>

<output>
  <restart  filename="C60-restart.xml" interval="10" />
  <position filename="C60-position.xyz" interval="4" />
</output>

</config>
```

Figure 2 : An example of XML formatted input file. Each item of conditions is specified by the tags <...> and </...>. In this example of input file, calculation consitions are set as the initial atomic structure file: C60.xml, integration algorithm for molecular dynamic calculation: velocity verlet, time step: 1.0fsec, total simulated time: 250fsec. Results of atomic structures are written in a file named "C60-position.xyz" every 4 steps, and restart file "C60-restart.xml" is updated every 10 steps.

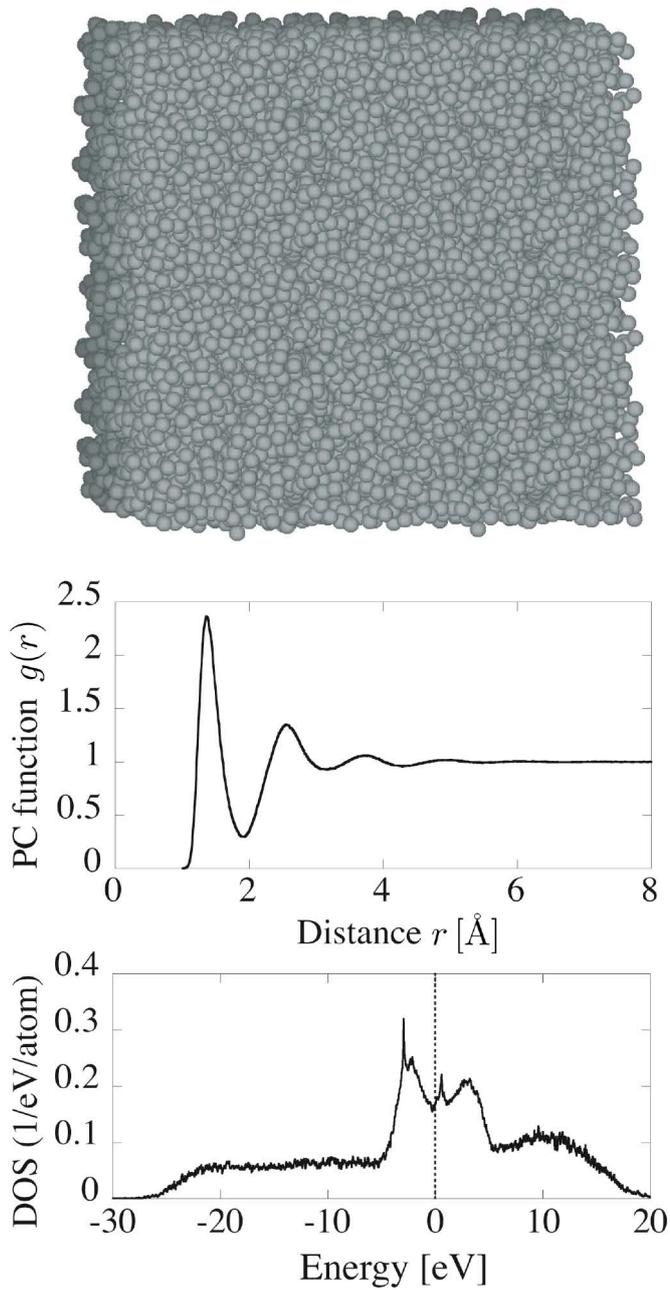

Figure 3 : A liquid carbon simulation with 13824 atoms is shown as an example of simulation result. The upper, middle and lower panels show atomic structure in the periodic simulation cell, pair correlation function, local density of states, respectively. [11]

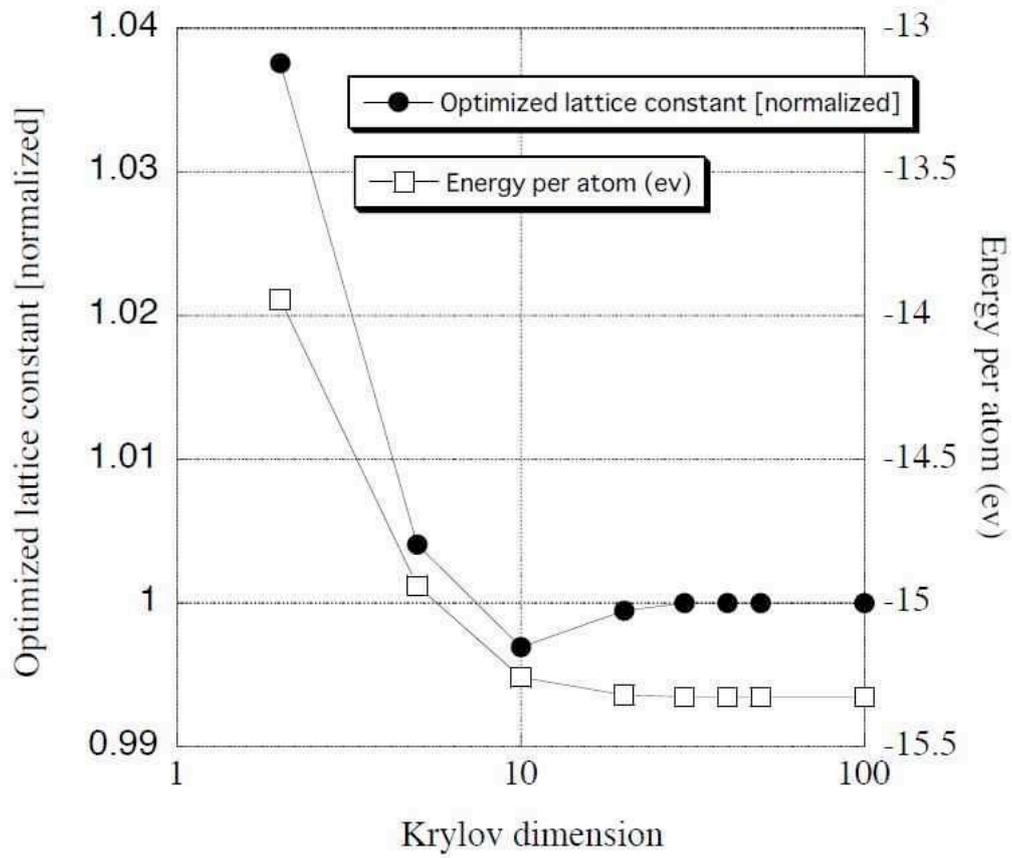

Figure 4 : The optimized lattice constant and the energy for bulk GaAs as a function of the Krylov dimension.